\documentclass[sigconf]{acmart}
\AtBeginDocument{%
  \providecommand\BibTeX{{%
    \normalfont B\kern-0.5em{\scshape i\kern-0.25em b}\kern-0.8em\TeX}}}

\setcopyright{acmcopyright}
\copyrightyear{2018}
\acmYear{2018}
\acmDOI{XXXXXXX.XXXXXXX}

\acmConference[Conference acronym 'XX]{Make sure to enter the correct
  conference title from your rights confirmation emai}{June 03--05,
  2018}{Woodstock, NY}
\acmPrice{15.00}
\acmISBN{978-1-4503-XXXX-X/18/06}

\usepackage{subcaption}
\usepackage{graphicx}
\usepackage{multirow}
\usepackage{amsmath}
\usepackage{comment}
\usepackage{footnote}
\usepackage{url}
\usepackage{xspace}
\usepackage{arydshln}
\usepackage{CJKutf8}
\usepackage{tcolorbox}
\usepackage{enumitem}
\usepackage{lipsum}
\usepackage{bbm}
\tcbuselibrary{skins}

\begin{document}

%\title{The Name of the Title is Hope}

% \title[Ranking User Satisfaction with Pretrained Language Models in Web Search 
% ]{Ranking User Satisfaction with Pretrained Language Models in Web Search 
% }
\title[Pretrained Language Model based Web Search Ranking: From Relevance to Satisfaction
]{Pretrained Language Model based Web Search Ranking: \\ From Relevance to Satisfaction
}
\author{Canjia Li$^\dagger$, Xiaoyang Wang$^\dagger$, Dongdong Li, Yiding Liu, Yu Lu, Shuaiqiang Wang, Zhicong Cheng, Simiu Gu, Dawei Yin$^*$}
\affiliation{
Baidu Inc. \city{Beijing} \country{China}\\
    \texttt{\{licanjia, wangxiaoyang06, lidongdong06, luyu06, wangshuaiqiang, chengzhicong01, gusimiu\}@baidu.com} \\
    \texttt{liuyiding.tanh@gmail.com}; \texttt{yindawei@acm.org} \\
    \vspace{5mm}
}
\thanks{$\dagger$ equal contribution.}
\thanks{$*$ Dawei Yin is the corresponding author.}

\renewcommand{\shortauthors}{Li, et al.}

\begin{abstract}
% Pretrained language models (PLMs) have become the de facto models in web search. 
Search engine plays a crucial role in satisfying users' diverse information needs.
Recently, Pretrained Language Models (PLMs) based text ranking models have achieved huge success in web search. 
However, many state-of-the-art text ranking approaches only focus on core relevance while ignoring other
dimensions that contribute to user satisfaction, e.g., document quality, recency, authority, etc. 
In this work, we focus on ranking user satisfaction rather than relevance in web search, and propose a PLM-based framework, namely SAT-Ranker, which comprehensively models different dimensions of user satisfaction in a unified manner.
In particular, we leverage the capacities of PLMs on both textual and numerical inputs, and apply a multi-field input that modularizes each dimension of user satisfaction as an input field. 
% Notably, most of the fields are presented in numbers, by which the numeracy skill of PLMs is unlocked to model semantic relevance and user satisfaction jointly.
% We call this new approach SAT-Ranker, as it aims to satisfy the diverse matching requirements rather than pinpointing core relevance matching.
Overall, SAT-Ranker is an effective, extensible, and data-centric framework that has huge potential for industrial applications. On rigorous offline and online experiments, SAT-Ranker obtains remarkable gains on various evaluation sets targeting different dimensions of user satisfaction. 
It is now fully deployed online to improve the usability of our search engine.
%When deployed online,SAT-Ranker achieve xx DCG gains, confirming the superiority of modeling diverse matching patterns.

\end{abstract}






\maketitle

\section{Introduction}
\label{sec:intro}

Search engines, such as Google and Bing, are useful tools for people to access rich information on the Internet. For web search, ranking relevance is the core problem~\cite{yin2016ranking},
% , aiming to rank the most relevant webpages for a given query on top of the search results. To model the relevance between the query and webpage, 
where extensive efforts from both industry and academia have been dedicated,
%~\cite{yin2016ranking,grbovic2018real,zhang2020towards,zou2021pre,lin2021pretrained}
such as text ranking approaches based on 
% on modeling relevance between query and webpages, 
% where the methodology evolves from 
% Early studies are mainly based on lexical matching (e.g., BM25~\cite{robertson2009probabilistic}), which aims to find exact keywords matching between queries and webpages.  
lexical matching (e.g., BM25~\cite{robertson2009probabilistic}) and semantic matching (e.g., DSSM~\cite{huang2013learning}, DRMM~\cite{guo2016deep}). More recently, Pretrained Language Models (PLMs) have shown promising performance for relevance modeling in web search \cite{zou2021pre,lin2021pretrained,fan2022pre,DBLP:conf/kdd/LiuLCSWCY21}.

\begin{figure}[t]
     \centering
     \begin{subfigure}[b]{0.23\textwidth}
         \centering
         \includegraphics[width=\textwidth]{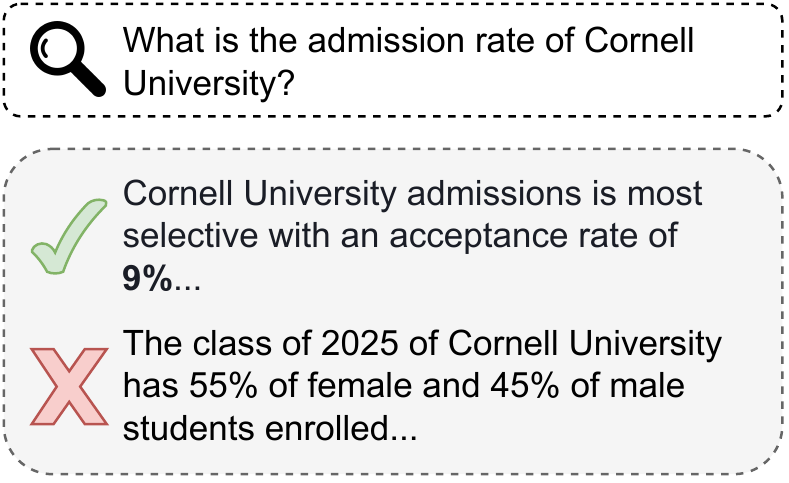}
         \caption{Relevance}
         \label{fig:five over x}
     \end{subfigure}
     \hfill
     \begin{subfigure}[b]{0.23\textwidth}
         \centering
         \includegraphics[width=\textwidth]{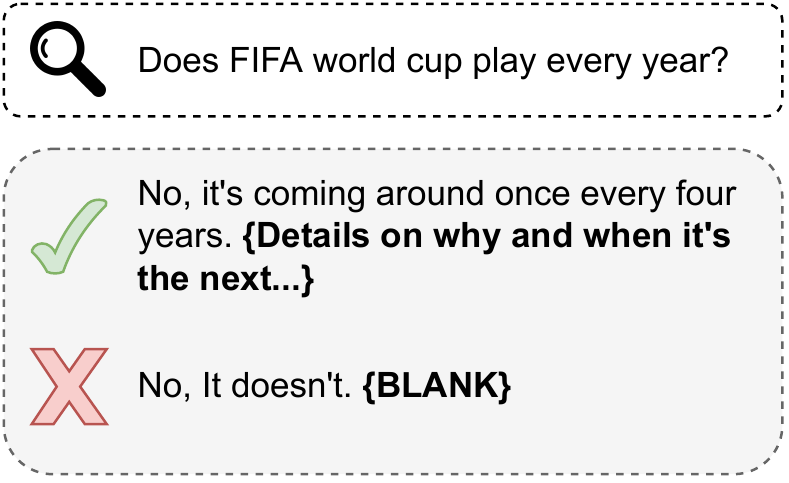}
         \caption{Quality}
         \label{fig:y equals x}
     \end{subfigure}
     
     \begin{subfigure}[b]{0.23\textwidth}
         \centering
         \includegraphics[width=\textwidth]{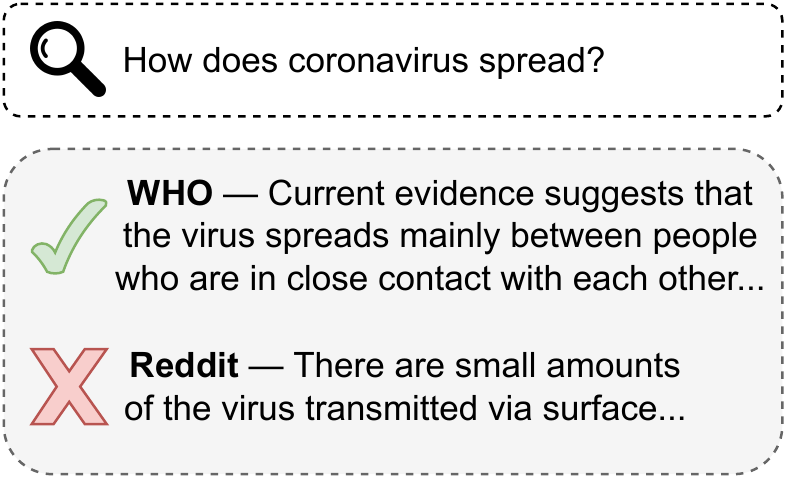}
         \caption{Authority}
         \label{fig:three sin x}
     \end{subfigure}
     \hfill
     \begin{subfigure}[b]{0.23\textwidth}
         \centering
         \includegraphics[width=\textwidth]{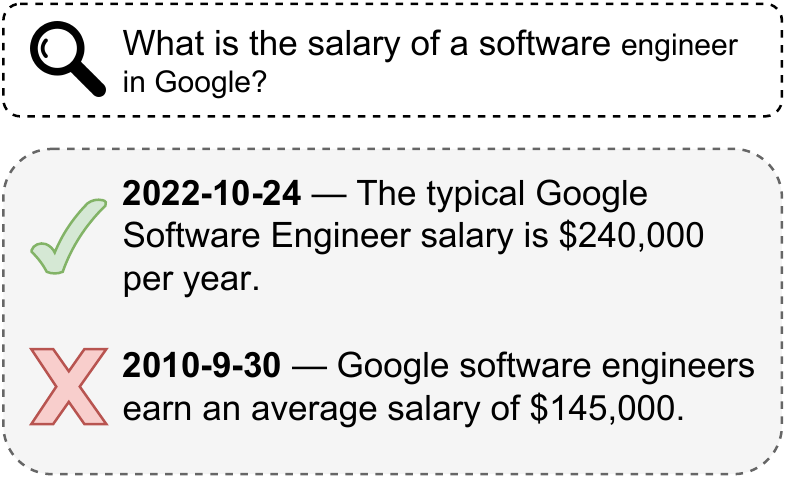}
         \caption{Recency}
         \label{fig:five over x}
     \end{subfigure}
        \caption{Aspects of user satisfaction. Each aspect is illustrated with a good and a bad result to the query. Note that in (b), (c), and (d), the good and bad results are relevant to the query, while the good results are more likely to satisfy users.
        }
        \label{fig:aspects}
\end{figure}
With the prosperity of Web 2.0, the information need of users has also become increasingly diverse, and an effective portal web search engine must satisfy a wide range of queries~\cite{dong2010time}.
In such case, \textit{relevance} alone is often insufficient to produce good rankings w.r.t. \textit{user satisfaction}. Previous work~\cite{dan2016measuring,mao2016does,yilmaz2014relevance} has found that relevance is not the equivalence of user satisfaction or usefulness of the search results. Incorporating authority~\cite{kleinberg1999authoritative,liu2008browserank,metzler2021rethinking} and recency~\cite{dong2010towards,yin2016ranking,dong2010time} of the search results can significantly improve user satisfaction. % with diverse information need.
Motivated by this, we can systematically formulate four critical dimensions for modeling diverse user satisfaction in real-world search engines as follows:
\begin{itemize}[leftmargin=5mm]
    \item \textbf{Relevance}. For all queries, relevance is the foundation of search ranking. The webpages returned to users must be relevant to the query. However, users' demands on relevance may also vary, where diverse relevance patterns (e.g., passage-level and document-level relevance) should be considered~\cite{fan2018modeling}.
    % both lexical relevance and semantic relevance need to be considered in different scenarios.
    \item \textbf{Quality}. For all queries, the quality of a webpage is also an important factor to satisfy users. The quality of a webpage represents its overall presentation and informativeness, i.e., whether users could effortlessly and effectively access useful information from its content. 
    % During the ranking stage, if two webpages have similar relevance levels with the query, the high-quality one should be ranked higher.
    \item \textbf{Authority}. For particular authority-sensitive queries, such as medical and legal inquiries, the information provided by authoritative websites, or the articles written by well-known authorities should be preferred by users. These search results are more likely to provide authentic information.
    % that contributes to the trustworthiness of search engines.
    % ~\cite{su2021beyond}
    \item \textbf{Recency}. For particular recency-sensitive queries, such as emergency events or frequently-updated information, users implicitly seek recently-published content, and those out-of-date information is valueless to the users. Typical recency-aware queries include news, stock, weather, etc.
    % updated policies, etc.
\end{itemize}
Figure~\ref{fig:aspects} gives illustrative examples. Although there might be other factors or taxonomies that could also be explored, such as users' locations~\cite{yin2016ranking} and search behaviors~\cite{dan2016measuring,grbovic2018real}, we do not enumerate them in this paper and stay focused on the 
% above-mentioned 
above four dimensions. 

For real-world search ranking, the major limitation of most PLM-based rankers is that they mainly focus on modeling relevance while ignoring users' diverse information needs. 
A handful of studies may notice that users' demands on relevance are diverse~\cite{gao2021complement,wang2021bert,fan2018modeling}.
% , and thus propose to combine lexical and semantic matching to capture relevance more comprehensively~\cite{gao2021complement,wang2021bert}. 
However, their methods are tailored for relevance modeling, which is not well aligned to user satisfaction. Therefore, it is appealing to develop a novel and practical solution that advances PLMs from ranking relevance to ranking satisfaction.
% , while the diversity of user satisfaction 
% There are few attempts that leverage PLMs to comprehend the diverse user behaviors in recommender systems~\cite{gu2020deep,cui2022m6,wang2022learning},
There are few notable attempts made to achieve a similar goal for 
recommender systems~\cite{gu2020deep,cui2022m6,wang2022learning},
% Advancing PLM to model multifaceted objective has achieved initial success in recommender system~\cite{gu2020deep,cui2022m6}, 
yet a unified framework for ranking diverse search satisfaction remains under-explored.

\vspace{2mm}
\noindent\textbf{Present work.}
% Previous PLM-based search models ~\cite{DBLP:series/synthesis/2021LinNY,DBLP:journals/corr/abs-1901-04085,DBLP:conf/kdd/LiuLCSWCY21,DBLP:conf/kdd/ZouZCMCWSCY21} fail to the aforementioned matching aspects as it pinpoints to semantic matching.
% Systematically combining all into PLMs is non-trivial yet.
% To bridge this gap, 
In this paper, we propose a generic Pretrained Language Model (PLM) based framework, namely 
% that modularizes various matching patterns in a single PLM.
\textbf{SAT-Ranker}, 
% {\bf ERNIE}~\cite{DBLP:conf/aaai/SunWLFTWW20} backbone (i.e., a PLM) 
to model diverse user satisfaction.
% diverse matching patterns
% Given a query, SAT-Ranker is informed by an off-the-shelf query analysis service whether a query asks for recent, or authentic information, or whether it is a navigational query.
In particular, the structure of SAT-Ranker is a standard transformer encoder~\cite{vaswani2017attention,devlin2018bert}, which could be easily deployed in our ranking system.
% , and pre-trained following our previous work~\cite{zou2021pre}. 
% To model user satisfaction 
% During fine-tuning stage, 
To empower SAT-Ranker for ranking satisfaction, we establish a multi-field input that covers the aforementioned dimensions in a unified manner. 
First, the raw text of the query, document title, and document summary are concatenated as the main body of the input, which provides semantic information for modeling the core relevance. 
Second, we leverage the numeracy ability of PLM and introduce several sets of numerical features to facilitate the model capability on lexical relevance, quality, authority, and recency. These features could provide rich information that can hardly be captured by the textual input.
% several lexical matching features as inputs, which complement lexical relevance for the model. 
% Furthermore, document-level features w.r.t. authority and recency are integrated, facilitating the model from ranking relevance to ranking comprehensive satisfaction. 
Third, we leverage the off-the-shelf query analysis service of our search system and incorporate query-level features to guide the training of SAT-Ranker for various kinds of queries, e.g., learning to focus on particular dimensions with respect to different demands.

% (usually in the form of numbers) are combined so that the matching distribution in each aspect is depicted.

In general, we leverage the capabilities of PLMs on comprehending both text and numbers~\cite{wallace2019nlp}, and modularize each dimension of user satisfaction as an input field in SAT-Ranker. As a result, we finalize an expressive, flexible, and practical framework for ranking user satisfaction in web search.
% Noteworthy, all fields are operated in numerical forms except the core relevance field.
% each matching aspect is modularized as a single input field in SAT-Ranker, ending up with a single, flexible, extensible, and unified matching model tailored for web search. 
Such a framework is appealing for industrial applications as 1) it moves beyond relevance modeling, 
% and systematically models diverse aspects for user satisfaction, 2) 
and leverages a powerful PLM to enable more expressive modeling of user satisfaction, 2) it is data-centric, without any modification of model architecture, which can be easily implemented and deployed in real-world productions, and 3) it is a unified framework that can be 
easily extended to support modeling other useful aspects w.r.t. user satisfaction.
% of both textual and numerical features in a unified manner, and 3) 
% \begin{itemize}[leftmargin=5mm]
%     \item It moves beyond one-dimensional relevance modeling, and systematically models diverse aspects for user satisfaction.
%     \item It unlocks the numerical reasoning ability of PLM for search. PLM is shown to have promising text understanding capabilities including numerical reasoning skill~\cite{DBLP:journals/corr/abs-2210-05075,DBLP:journals/corr/abs-2205-06733}. They show superior performance on tabular data~\cite{DBLP:journals/corr/abs-2209-08060,DBLP:journals/corr/abs-2207-03208,DBLP:conf/naacl/IidaTMI21,DBLP:journals/corr/abs-2012-06678}.
%     Such ability can be transferred to the ranking tasks without extra effort. 
%     \item The framework is unified, flexible, and extensible. It enables deeper interaction of the matching patterns from the bottom.
%     % One can easily extend the input into more aspects that benefit user satisfaction.
%     \item The framework is data-centric, without any modification of model architecture, which can be easily implemented and deployed in real-world productions empowered by existing deep learning architectures.
% \end{itemize}
% The advantages of such input are further studied in Section x.
Overall, our contributions are threefold:
\begin{itemize}[leftmargin=5mm]
    \item We move beyond conventional relevance-oriented ranking, and formulate the problem of ranking user satisfaction, which is critical for empowering real-world search engines.
    % The definition of diverse matching aspects of PLM for search.
    % \item A formalization of modeling diverse matching patterns using a pretrained ERNIE model.
    \item We propose a unified framework, namely SAT-Ranker, to model user satisfaction.
    It leverages the expressiveness of PLM for both textual and numerical features to comprehensively model different aspects of user satisfaction.
    \item We conduct rigorous offline and online evaluations to demonstrate the effectiveness of SAT-Ranker. 
    % SAT-Ranker achieves x nDCG@4 gains on an in-house ranking dataset and xx nDCG@4 gains on the online evaluation.
    % We conduct extensive offline and online experiments to validate the effectiveness of the retrieval system. 
    The results show that SAT-Ranker can obtain remarkable gains on various evaluation sets targeting different aspects of user satisfaction. It is now fully deployed online to improve our search engine.
    % and has significantly improved user satisfaction with our search engine.
    % Offline and online evaluation of the proposed method in industry
\end{itemize}

\begin{figure*}
     \centering
     \begin{subfigure}[b]{0.46\textwidth}
         \centering
         \includegraphics[width=\textwidth]{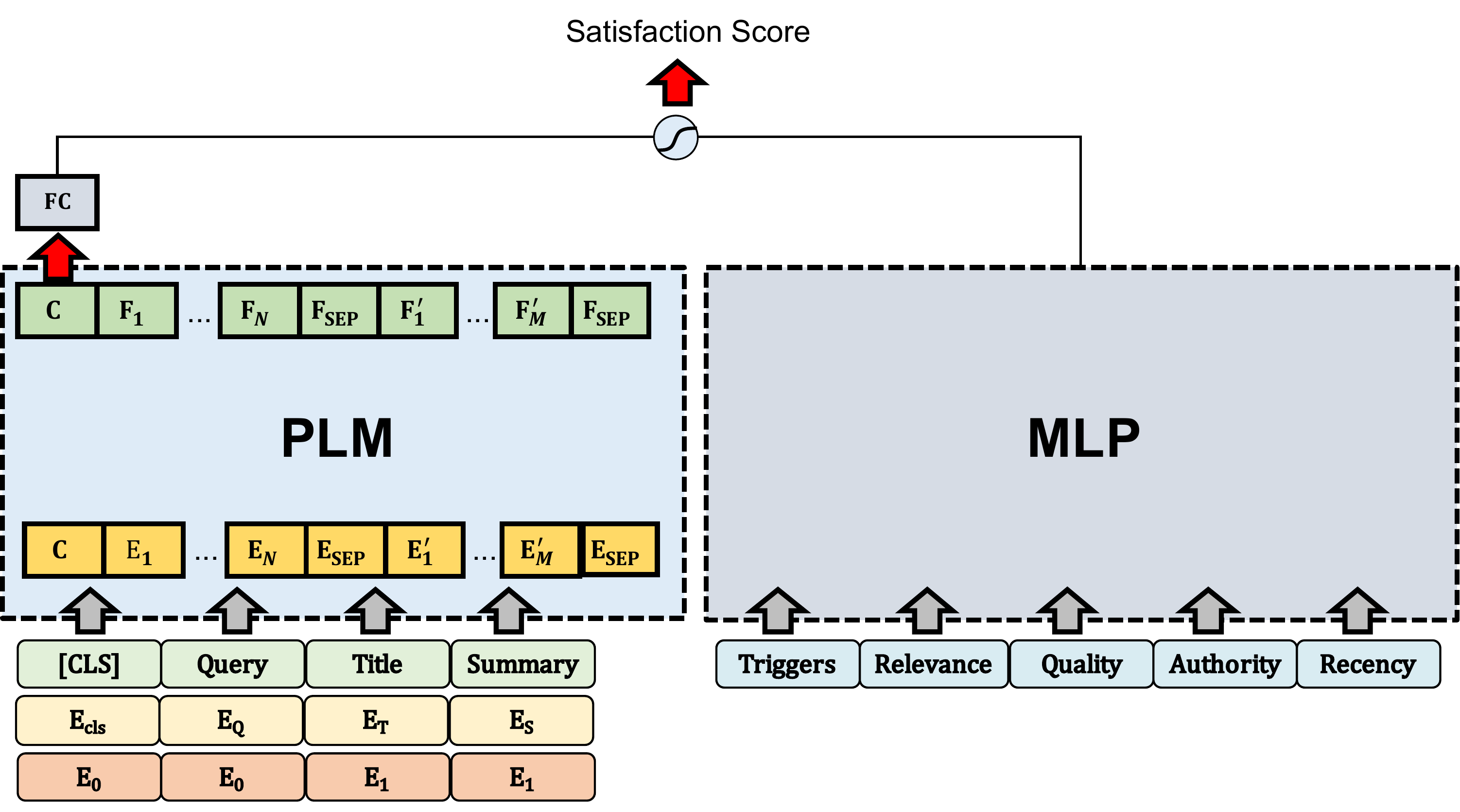}
         \caption{SAT-Ranker$_{mlp}$. The textual input and numerical input are decoupled into a PLM-based text encoder and a MLP network.}
         \label{fig:base}
     \end{subfigure}
     \hfill
     \begin{subfigure}[b]{0.46\textwidth}
         \centering
         \includegraphics[width=\textwidth]{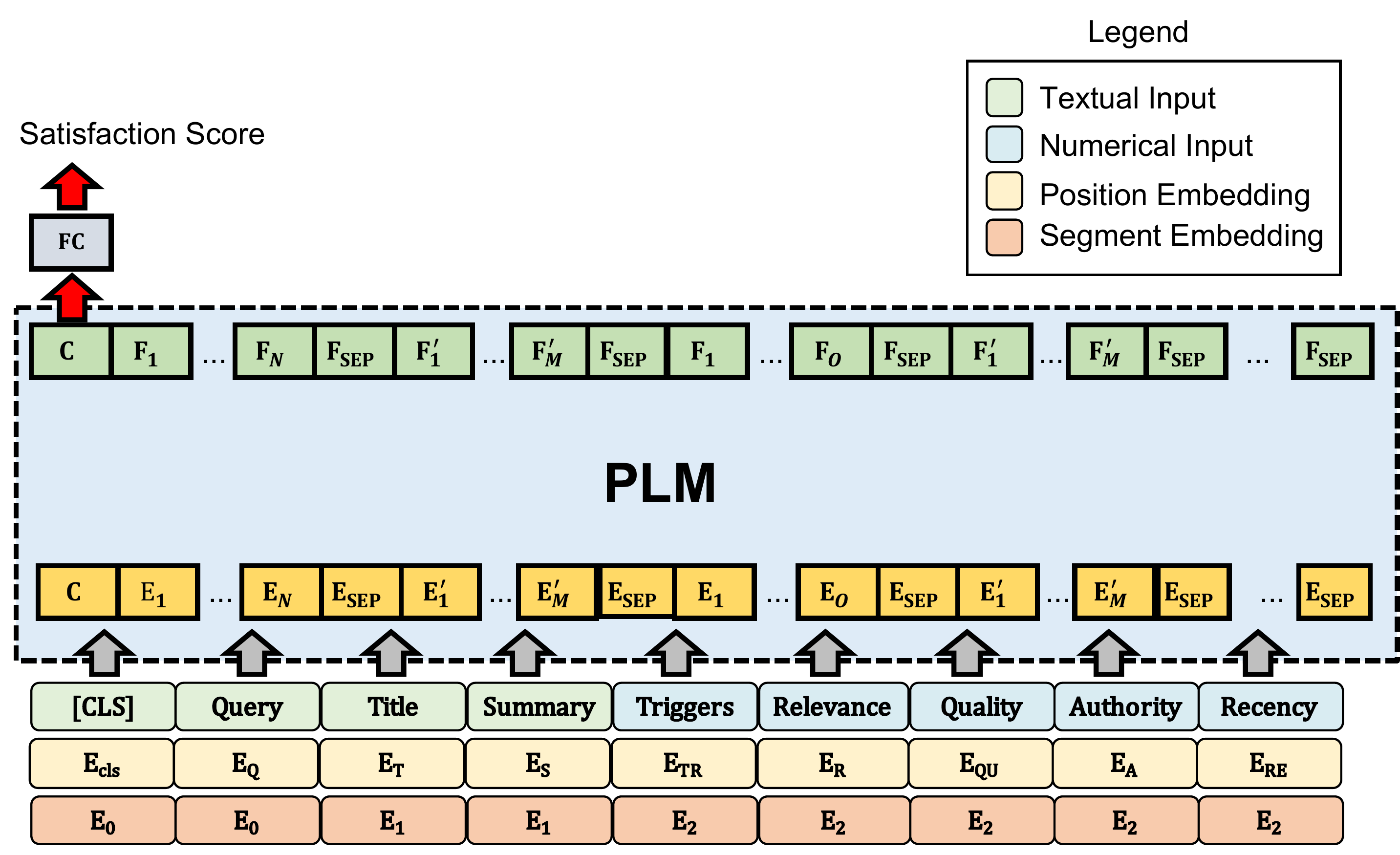}
         \caption{SAT-Ranker. Both textual input and numerical input are jointly learned from the bottom.}
         \label{fig:stgy}
     \end{subfigure}
        \caption{Overall frameworks of SAT-Ranker$_{mlp}$ and SAT-Ranker.}
        \label{fig:framework}
\end{figure*}
% \section{SAT-Ranker}
\section{Framework}
\label{sec:method}

In this section, we first present the task formulation of modeling 
% relevance beyond semantic matching
user satisfaction, and then propose our framework, namely SAT-Ranker, which considers the diverse information needs of users
% to incorporate diverse matching patterns 
to handle this task.

\subsection{Task Formulation}
\label{sec:task}
\noindent\textbf{Ranking}. A search engine is usually composed of retrieval and ranking modules and our focus is on the ranking stage. 
% We use the terms ``ranking'' and ``ranking'' interchangeably in the rest of this paper.
% SAT-Ranker is a ranker embedded in the ranking module that ranks the top $k$ documents retrieved by the retriever.
In particular, we apply the \emph{score-and-sort} setting, where the ranker is considered as a pointwise scoring function $s(q,d;\Theta)$ for a query $q$ and a candidate document $d$,
% with $n$ candidate documents (i.e., webpages),
and $\Theta$ denotes the trainable model parameters. The ranker $s(q,d;\Theta)$ is trained to map the input features of $q$ and its $n$ candidate documents to $n$ real number scores (denoted as $\mathbf{s}_q\in\mathbb{R}^n$), where the training objective is to minimize the loss:
\begin{equation}
    \mathcal{L}(\Theta) = \frac{1}{|Q|}\sum_{q\in Q}l(\mathbf{y}_q, \mathbf{s}_q),
\end{equation}
where $Q$ is the set of all training queries, and $l(\cdot)$ is a ranking loss (e.g., pairwise loss~\cite{burges2010ranknet} and listwise loss~\cite{qin2010general}) for a single query that measures the consistency between $\mathbf{s}_q$ and the ground truth $\mathbf{y}_q$. 

\noindent\textbf{Labels}. Most of the existing rankers are trained and evaluated on data with annotated relevance~\cite{nguyen2016ms}, which cannot be applied to model user satisfaction. Instead, we employ a user satisfaction dataset to supervise the ranker, where $\mathbf{y}_q$ is labeled with more comprehensive user satisfaction. Specifically, each query-document pair is graded in 0-4 ratings, indicating \{bad, fair, good, excellent, perfect\}, respectively, in terms of overall user satisfaction.

\noindent\textbf{Features}. Following previous study~\cite{zou2021pre}, the main input of our ranker is composed of multiple text fields, i.e., query, document title, and document summary~\cite{DBLP:journals/corr/abs-2210-08809}. 
The length of the document summary can be varied based on the latency and computational cost during online serving. 
While such textual input might be sufficient for modeling relevance, it can hardly reveal other dimensions of user satisfaction. To this end, we further include more numerical and textual attributes associated with $q$ and $d$ as new input fields. In particular, we consider three types of attributes, i.e., query attributes $A_q$, document attributes $A_d$ and query-document interaction attributes $A_{qd}$, which intend to cover all four dimensions of user satisfaction. As such, the ranker can be defined as
% Most of the existing rankers either 

% For each document $d$, 

% Given a query $q$ and a candidate document $d$, a ranker aims at 

% SAT-Ranker generates a relevance score as:
\begin{equation}\label{eq:ranker}
    s(q, d;\Theta) = f(text_q, text_d, A_q, A_d, A_{qd};\Theta).
\end{equation}
In the following, we elaborate our proposed SAT-Ranker framework as $f$ for modeling user satisfaction.
% Where $f_{\theta}$ is an ERNIE backbone with parameter $\theta$.
% SAT-Ranker models the diverse matching patterns through $A_q$, $A_d$, and $A_{qd}$, which are generated by off-the-shell services in the search engine.
% Without these attributes, SAT-Ranker reduces to a matching model solely on texts~\cite{zou2021pre,DBLP:journals/corr/abs-1901-04085}.

% \subsection{ERNIE Ranker}

\subsection{The SAT-Ranker Framework}
Given the task formulation, the key to ranking user satisfaction is to design a practical framework that can 1) expressively model the basic relevance based on the textual inputs, and 2) flexibly incorporate heterogeneous attributes to capture different dimensions of user satisfaction. 
Next, we introduce the text-based ranker and our proposed SAT-Ranker method that address the two concerns.

\noindent\textbf{Text-based ranker}. 
To model the basic semantic relevance, we apply a PLM~\cite{DBLP:conf/aaai/SunWLFTWW20} as the backbone for our ranker. Specifically, we adopt a transformer-based pretrained language model that can be further finetuned for search ranking. We use the base version of the PLM that has 12 layers, 12 heads, and 768 hidden dimensions, which results in a total of 110M parameters. The expressiveness of PLM is beneficial for understanding the semantics underlying the textual inputs of query and document. Following previous study~\cite{zou2021pre}, a text-based ranker is trained with a multi-stage paradigm.
% It is experimented with and verified to be effective for search ranking~\cite{zou2021pre}.
% \noindent\textbf{Multi-field input}. 

To extend the basic text-based ranker for modeling user satisfaction, we develop two frameworks that are flexible and extendable to incorporate rich attributes, namely SAT-Ranker$_{mlp}$ and SAT-Ranker. The overall architectures are shown in Figure \ref{fig:framework}. 
% Specifically, 

\noindent\textbf{SAT-Ranker$_{mlp}$} adopts a hybrid structure that consists of the PLM-based encoder and an extra multi-layer perceptron, as inspired by wide\&deep structure~\cite{DBLP:conf/kdd/ZhouZSFZMYJLG18}. In particular, the PLM and MLP modules are employed to process textual and numerical features, respectively. As shown in Figure \ref{fig:base}, the numerical features are carefully designed to cover all the dimensions of user satisfaction, i.e., relevance, quality, authority, and recency, especially for those aspects that can hardly be revealed by the textual features. By doing this, the main semantic relevance and other dimensions are separately modeled 
% by two modules, i.e., the ERNIE backbone and the MLP, respectively 
in SAT-Ranker$_{mlp}$, and the final ranking score is the summation of the scores predicted by the two modules\footnote{
% \revision{
We have also tried other implementations like injecting the final score or the final \texttt{[CLS]} representation from PLM to the MLP network, but do not get better results.
% }
}. 
% To train SAT-Ranker$_{mlp}$, we first optimize ERNIE module to fit the satisfaction 

\noindent\textbf{SAT-Ranker} is a more generic framework of unifying textual and numerical features for modeling user satisfaction. Particularly, it converts all the numerical features to token embeddings, which are concatenated with the word embeddings of the textual inputs. Compared with SAT-Ranker$_{mlp}$, SAT-Ranker is a more favorable solution as 
% 1) it perceives the numerical features with more expressive transformers, instead of simple MLP, and 2) it 
can capture complicated relationships between different dimensions towards final user satisfaction. In contrast, SAT-Ranker$_{mlp}$ hinders the deep interaction between numerical features and textual features and thus is less capable. In the next section, we depict the detailed design of the input features used in SAT-Ranker.

\section{Features of User Satisfaction}
In this section, we present the detailed design of input features that allows SAT-Ranker to comprehend user satisfaction. Despite the main textual inputs, i.e., the texts of query, document title, and document summary, we further introduce several numerical features of relevance, quality, authority, and recency in the following.
% We elaborate on both textual and numerical features integrated into SAT-Ranker.
% We introduce several aspects that contribute to user satisfaction, including semantics, recency, authority, navigation, and quality.

% \subsubsection{Relevance}

% \noindent{\bf Relevance.} 
\subsection{Relevance} 
% In search ranking, relevance is the core of user satisfaction, where 
% Conventional 
Although PLM-based rankers have already achieved remarkable performance in modeling semantic relevance,
% PLM-based rankers have achieved remarkable performance on relevance ranking, while the computational cost and online latency are very high if all the document content is fed into the model.
% . However, their computational cost scales quadratically to the text length, which is infeasible for real-world search engines serve massive long documents with very low latency. 
% Former experience~\cite{zou2021pre} indicates that query-aware document summary is a promising way of reconciling efficiency and effectiveness. 
% However, 
empirical studies~\cite{wu2019investigating,fan2018modeling,DBLP:journals/corr/abs-2008-09093} further highlight the importance of full-document signals, which are neglected by the short document summary applied by PLM-based methods~\cite{zou2021pre,DBLP:journals/corr/abs-2210-08809}. 
Therefore, we leverage several numerical features to represent full-document relevance signals, namely $A_{qd}^{rel}$, which supplement diverse relevance patterns neglected by document summary, including
% Despite its overall performance, such a method has a clear drawback in modeling lexical matching. For example, 
% A majority of neural IR models~\cite{DBLP:journals/ipm/GuoFPYAZWCC20} and PLM-based IR models~\cite{DBLP:series/synthesis/2021LinNY} focus on relevance matching.
% Regarding long-form document semantic matching, previous works either operate on sentences~\cite{DBLP:conf/emnlp/YilmazWYZL19} or passages~\cite{DBLP:journals/corr/abs-1901-04085} or full-text~\cite{DBLP:journals/corr/abs-2008-09093}.
% In industry, using either passages or full text is intractable due to compute budget.
% As a trade-off, we use a query-aware sentence extraction technique to extract the two most relevant sentences as a proxy to the document~\cite{zou2021pre}.
% The extraction is based on the overlap between the query and sentence.
% Readers are referred to~\citet{zou2021pre} for more details.

% Under the SAT-Ranker framework, 
% In particular, 
% we consider the lexical matching between query and full document content, which is very efficient to compute online.
% patterns into account, which complements the two extracted sentences.
% we design a set of full-text matching features, denoted as $A_{qd}^{rel}$,
% to measure the diverse relevance patterns, including 
% 1) term coverage, 2) term proximity, and 3) term continuity:
\begin{itemize}[leftmargin=5mm]
    \item \textbf{Term coverage}. It measures the ratio of query terms that can be matched in the document, or we can also use existing methods like BM25~\cite{robertson2009probabilistic};
    \item \textbf{Term proximity}. It measures the distance of matched terms in the document for each query bi-gram, considering that term matches that are closely distributed in a document are a more reliable relevance signal than those randomly scattered all over the document;
    \item \textbf{Term continuity}. It measures the longest continuous matching between the document content and the query, which is a strong signal of relevance.
\end{itemize}
Note that we only give some loose definitions and omit the details, as the framework is flexible to support any specific implementation. These features are very efficient to be computed online compared with the PLM backbone, and the computational overhead could be neglected. Overall, these full-text relevance features are able to complement text-based rankers that rely on document summaries. 
% Besides, SAT-Ranker allows end-to-end modeling of both lexical and semantic relevance, which is 

% matched segment.
% These features are capable of depicting the matching distribution of the document.
% Given the scalar matching features, one can interpolate scores for model ensemble~\cite{DBLP:conf/emnlp/YilmazWYZL19,DBLP:journals/corr/abs-2010-00200}.
% By injecting these matching features into a specific field, our SAT-Ranker framework opens up the opportunity for text-based features and numerical features (e.g., $A_{qd}^{sem}$) to interact in an end-to-end fashion.
% The novel simple interaction also alleviates the burden of tuning the hyper-parameters for the score ensemble.

\subsection{Quality} 
% \noindent{\bf Quality.} 
The quality of web documents can hardly be revealed by conventional PLM-based rankers, as the input texts, i.e., query, title, and document summary, mainly reflect relevance. To this end, we design two types of quality features $A_d^{qua}$ that could be involved in SAT-Ranker:
% Boosting highly relevant documents is challenging as documents are represented in various forms.
\begin{itemize}[leftmargin=5mm]
    \item \textbf{Statistical features}. These are valuable information sources of document quality. For example, we can apply document length, the number of pictures and videos, and the number of ads on the webpage as signals to reveal webpage quality.
    \item \textbf{DNN features}. Unlike relevance features that need to be computed on-the-fly with a specific query, document quality is query-agnostic and thus can be pre-computed offline and cached. Therefore, it is feasible to apply other expressive DNN models to extract insightful quality information from the full webpage, such as whether the content is well structured and presented.
\end{itemize}
% We model the document quality $A_d^{qua}$ through document length and users' dwell time.
% As a characteristic of the document, document length is a provider for the richness of the document.
% In conjunction with document length, dwell time reveals how much time a user would like to spend on a document.
% Intuitively, a lengthy document with a short dwell time is likely to be under quality.
% Putting it all together, SAT-Ranker can make a more moderate judgment on document quality.
In real scenarios, the statistical features are easier to obtain and apply, while DNN features usually require extra effort to design the objective and acquire training data w.r.t. webpage quality.

\subsection{Authority}
% \noindent{\bf Authority.}
% As web content emerged from random users, content reliability differs according to the author or site of the document. 
% Identifying relevant documents with authentic producers is challenging since the data is enormous and sparse.
%Two different cases are library search and academia search, where the items are from authentic sources.
% To tackle the challenge, SAT-Ranker is informed by the off-the-shelf service whether a query demands authentic content, coined as $A_q^{auth}$.
% $A_q^{auth}$ is categorized into three levels: low (0), medium (1), and high (2).
% For the document side features $A_d^{auth}$, 
% We also weigh a producer authority that involves statistics features (click, like, etc) on top of PageRank (TO BE CONFIRMED).
% We also compute the coverage of anchors to the query, namely $A_{qd}^{auth}$, weighted by the anchor's PageRank value.
Authority can also be represented by query-agnostic features of a document (denoted as $A_{d}^{auth}$), which could be pre-computed offline. In particular, there are three dimensions that are critical for modeling document authority:
\begin{itemize}[leftmargin=5mm]
    \item \textbf{Site}. The intuition is that the information published on an authoritative site (e.g., a government website) should more trustworthy and preferred by users. Specifically, we can tag the authoritative sites as a binary indicator, e.g., whether the content source is a government website. Alternatively, we can simply put the site name at a feature in SAT-Ranker, as our model can naturally process textual input.
    \item \textbf{Author}. As the web is flooded with user-generated content, such as Twitter and YouTube, it is also important to identify authoritative authors. Some simple statistics could be useful, such as the number of followers or subscriptions and whether the author has official verification on the platform, etc. These authors are also more likely to present trustworthy content.
    \item \textbf{Hyperlink}. In addition to the candidate document itself, one can also leverage its hyperlinks to measure the authority, such as PageRank~\cite{page1999pagerank}.
\end{itemize}
It is worth noting that we only consider the authority of the information source, while the correctness of the information cannot be guaranteed. In fact, toxic information posed by authorities (e.g., domain experts) is even more harmful to users. We rely on other fact verification systems to resolve this issue.

% \noindent{\bf Recency.}
\subsection{Recency}
Compared with other dimensions, recency is more straightforward to be modeled. Specifically, we consider the post time of a document as its recency feature. 
However, directly using the date or time as the feature is sub-optimal,  
% Note that the post time should be normalized before feeding to SAT-Ranker. 
% This is because that
as web content is emerging and expiring at a very high speed. A document could be considered with high recency at the time of training but becomes out-of-date when it is being presented to users. To this end, we apply a listwise normalization for the post-time during model training and online inference. In particular, for the candidate documents of a given query, we normalize their absolute post time to a relative post time, where the more recent documents have higher feature values. We denote the normalized recency features as $A_{d}^{rec}$.
% our goal is to rank the recency of a list of documents, we can normalize the post time in a listwise manner, which finalizes a relative recency score in the list.
% Therefore, we introduce two normalized post time as the recency features (denoted as $A_{d}^{fre}$):
% \begin{itemize}[leftmargin=5mm]
%     \item \textbf{Current time - post time}. We can use the number of hours/days away from the current time as a feature to represent recency.
%     \item \textbf{Listwise normalization}. Alternatively, if our goal is just to rank the recency of a list of documents, we can normalize the post time in a listwise manner, which finalizes a relative recency score in the list.
% \end{itemize}
% As an increasing number of documents emerged on the web, finding the newest relevant documents is challenging.

% recency is another demanding matching aspect for search engines.
% SAT-Ranker includes query-side feature $A_q^{fre}$ and document-side feature $A_d^{fre}$.
% A query recency level is given by: low (0), medium (1), and high (2). 
% By that, SAT-Ranker is aware of how timely and demanding the query is.
% On the document side, we include the publishing date of the documents.
% The dates are min-max normalized in a listwise manner.
% Therefore, SAT-Ranker knows the relative relationship of the documents rather than using the raw date. 
% We conduct an ablation study on how different forms of dates affect the model effectiveness in section xx.

\begin{table}[]
\centering
\caption{
Summary of features used for user satisfaction. 
% Semantic matching approaches are fundamentally based on texts.
SAT-Ranker facilitates user satisfaction via incorporating query attributes $A_q$, document attributes $A_d$, and query-document interaction attributes $A_{qd}$. The triggers $A_{q}^{trig}$ are also included.
}
\label{tab:summary}
\begin{tabular}{lccc} \toprule
{\bf Aspect}     & {\bf Query} & {\bf Document} &  {\bf Query-document} \\ \hline
Relevance   & text$_q$    & text$_d$      & $A_{qd}^{rel}$             \\
Quality    &      & $A_{d}^{qua}$       &               \\
Authority  &  $A_{q}^{trig}$   & $A_{d}^{auth}$       &             \\
Recency  &  $A_{q}^{trig}$   & $A_{d}^{rec}$        &               \\
% Triggers    &   $A_{q}^{trig}$   &        &               \\
\bottomrule
\end{tabular}

\end{table}

% {\bf Navigation.}
% \citet{li2020mpii} showed that pretrained language models suffer from navigational queries and suggest that adopting exact matching signals can be more reliable.
% They suggest using multiple document fields like URL, analogously to NRM-F~\cite{DBLP:conf/wsdm/ZamaniMSCT18}.
% For this aspect, we use an indicator, namely $A_q^{nav}$, to inform SAT-Ranker whether it is a navigational query.
% We adopt a pretrained sequence-to-sequence generative model to predict the site of a given query.
% To prepare with site predictive ability, the generative model is pretrained on a search log for query and site understanding.
% We then measure whether the document URL matches the predictive site, resulting in a query-document matching signal $A_{qd}^{nav}$.

\subsection{Triggering Authority and Recency}
As mentioned in Section~\ref{sec:intro}, relevance and quality are important dimensions for \emph{all queries}, while authority and recency are critical for particular \emph{authority-sensitive} and \emph{recency-sensitive} queries, respectively. However, such kinds of queries only take a small portion of all queries. For a given query, it is difficult for the model to determine whether authority and recency should be considered in the ranking. To tackle this challenge, we further introduce two types of query features (denoted as $A_{q}^{trig}$) that help the model to comprehend authority and recency when necessary:
\begin{itemize}[leftmargin=5mm]
    \item \textbf{Explicit triggers}. We can build an off-the-shelf query analysis that outputs whether a query is authority-sensitive, recency-sensitive, or both. The binary indicators could be included as features in SAT-Ranker.
    \item \textbf{Implicit triggers}. Moreover, other query analysis services in the search system could also be used as implicit triggers, such as domain of a query. Intuitively, some domains (e.g., healthcare and law) require more information from professionals and authorities, while some other domains (e.g., news and events) require high recency.
\end{itemize}

\subsection{Final Input Schema}
% The diverse matching patterns are summarized in 
Table~\ref{tab:summary} summarizes all the features included in SAT-Ranker, which moves beyond semantic relevance modeling and facilitates user satisfaction modeling.
% from the view of diverse matching aspects.
Specifically, the final input schema of SAT-Ranker for a query $q$ and a document $d$ is formalized as
\begin{tcolorbox}[width=3.3in,
                  standard jigsaw,
                  opacityback=0,
                  %%enhanced,
                  boxrule=0pt, 
                  frame empty,
                  interior hidden,
                  boxsep=0pt,
                  left=10pt,
                  % right=4pt,
                  % top=4pt,
                  ]%%
  
  \texttt{[CLS]} Q \texttt{[SEP]} 
T \texttt{[SEP]} 
S \texttt{[SEP]}

 $A_{q}^{tri}$ \texttt{[SEP]}
 $A_{qd}^{rel}$ \texttt{[SEP]}
$A_{d}^{qua}$ \texttt{[SEP]}
 $A_{d}^{auth}$ \texttt{[SEP]} 
 $A_{d}^{rec}$ \texttt{[SEP]} 
\end{tcolorbox}
\noindent The output representation of \texttt{[CLS]}
% , regarded as a final relevance feature, 
is fed into a feed-forward layer to generate the final score, as defined in Eq. (\ref{eq:ranker}). Note that we do not use any template or schema to prompt the numerical features, where the discussion and experiment of this could be found in Section \ref{dis:prompt} and \ref{sec:prompt_type}, respectively.

\begin{figure}
    \centering
    \includegraphics[width=0.46\textwidth, height=0.3\textwidth]{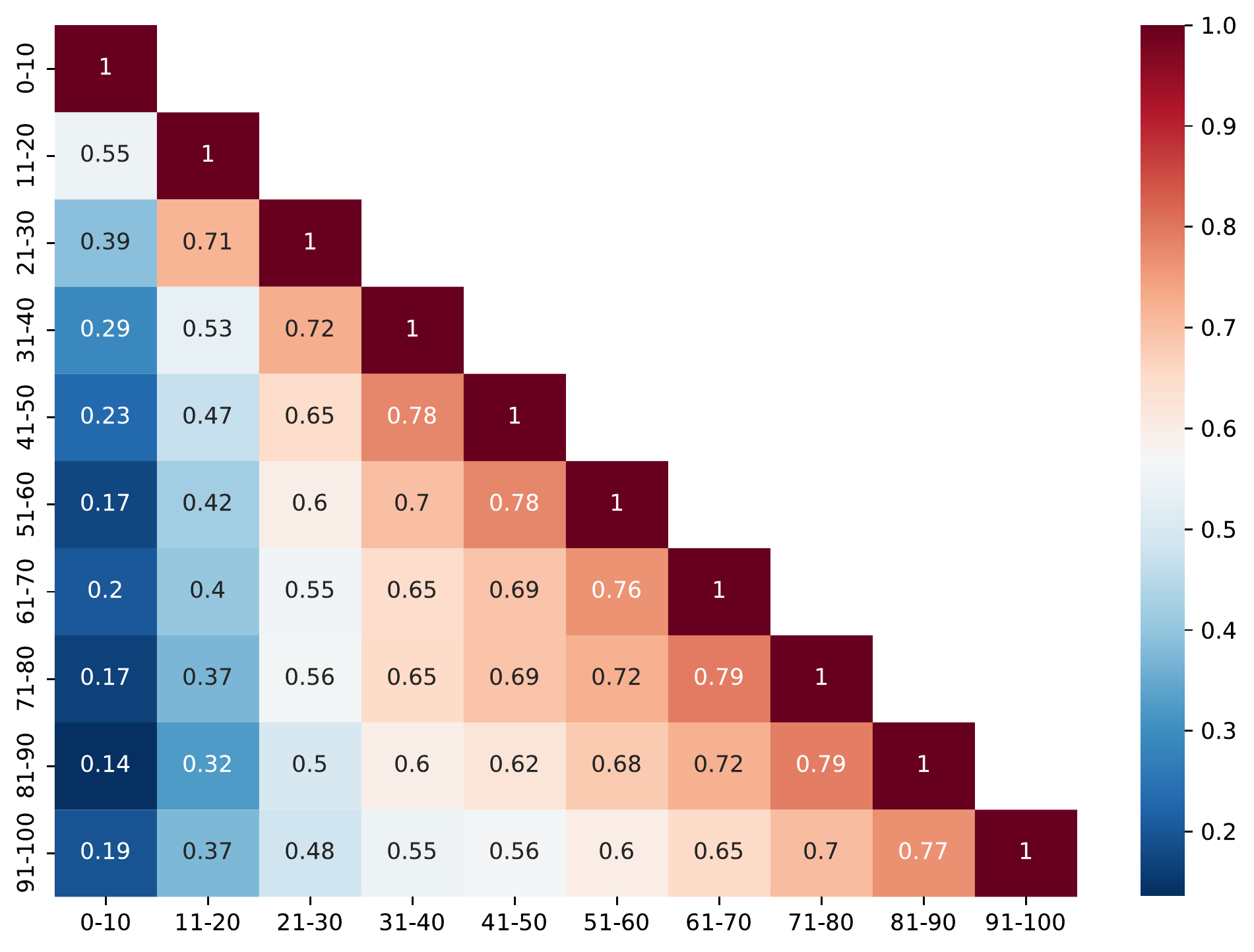}
    \caption{Pairwise cosine similarity of digital centroids. Each centroid is obtained by averaging the token embeddings in the corresponding range. For example, the cosine similarity between centroid(0-10) and centroid(91-100) is 0.19. 
    % This shows that PLM encodes the relative magnitude for numbers.
    }
    \label{fig:numeracy}
\end{figure}

\section{Discussion}
% \noindent\textbf{Numeracy}. 
\subsection{Numeracy in PLM}
% ~\citet{wallace2019nlp} propose several synthetic tasks to evaluate numeracy including {\it list maximum, decoding, and addition}. 
% We find out that SAT-Ranker needs to leverage the numerical capacity of the PLM (i.e., a pre-trained ERNIE) to model user satisfaction. Its performance degenerates dramatically when we replace the pretrained number embeddings with randomly initialized number embeddings.
% before fine-tuning SAT-Ranker, fine-grained information about number magnitude and order.
% We observe that 
The numerical capability of PLM might be the key to integrating numerical features in SAT-Ranker. Previous work~~\cite{wallace2019nlp} reveals that fine-grained information about number magnitude and order is naturally encoded in the pretrained embeddings, which is in line with our finding.
% We also find out that the pre-trained ERNIE is able to tell the magnitude of numbers.
% investigate the pre-trained number embeddings of ERNIE by measuring the distance between their centroids.
Specifically, Figure~\ref{fig:numeracy} demonstrates the cosine similarity between different ranges of numerical embeddings in the pre-trained model, and we can observe that the closer magnitude has a larger cosine similarity.
It reveals that our PLM indeed encodes the magnitude for numbers. 
% which is in line with the previous finding~\cite{wallace2019nlp}.
Note that the numeracy skill emerged during the pre-training stage, instead of during the target fine-tuning stage. 
In our pilot study, we also observed that the performance of SAT-Ranker degenerates dramatically when we replace the pretrained number embeddings with randomly initialized number embeddings.
These observations motivate us to devise a framework that leverages off-the-shelf multi-field features presented in numbers to model user satisfaction. To the best of our knowledge, this is the first work that leverages the numeracy skill of PLM to model user satisfaction in web search. However, we do not conduct rigorous experiments to quantify the influence of numeracy, as it is orthogonal to this paper. An in-depth investigation of how the numeracy of PLM affects the downstream task is left for further work.
% A very recent work~\cite{askari2023injecting} proposes to inject the BM25 score, which yields significant improvements over a single cross-encoder or the linear interpolation ensembles of BM25 and cross-encoder.
% demonstrate PLM's numeracy skill
% In SAT-Ranker, the numeracy skill is leveraged to model user satisfaction by using numbers up to 200.
%For the table parsing task, TABERT combines the textual input with schema input in a tabular to improve numerical representation~\cite{yin2020tabert}, which shares some similarities with our work.
%Differently, our work operates in a null prompt way rather than using schema~\cite{DBLP:journals/corr/abs-2203-00902}.

% \noindent\textbf{Tree-based methods}. 

% \noindent\textbf{Prompt}. 
\subsection{Prompt vs. Null Prompt}\label{dis:prompt}
It is worth noting that the numerical inputs are injected to SAT-Ranker and operate in a null prompt way~\cite{DBLP:journals/corr/abs-2203-00902}. We have also considered using schema prompts for different numerical fields (e.g., as an example shown in Table \ref{tab:prompt}), while the results turn out to be on par. Detailed experimental results can be found in Section~\ref{sec:prompt_type}. We do not spend much effort designing the template, and thus there is still a potential for exploring prompting methods for SAT-Ranker. Besides, another limitation of prompting is that it would significantly increase the length of the input. As a result, the online inference latency would be very high when the number of fields or features is very large. This would be an issue as ranking models demands high efficiency online.
% We study the effectiveness of different types of prompts 

% In summary, the advantages of SAT-Ranker over vallina ERNIE are:
% \begin{enumerate}[leftmargin=5mm]
%     \item Besides text-based semantic matching features, it modularizes several numerical feature fields as proxies to Lexicon matching, recency, Authority, and Quality.
%     \item The numerical feature fields use null prompts~\cite{DBLP:journals/corr/abs-2203-00902}, dramatically shortening the sequence length which is critical for online serving.
%     \item The numerical feature fields are of high positional invariance~\cite{geva2020injecting}. They are concatenated immediately after the textual inputs and do not need to be constrained at fixed positions.
%     \item The embeddings of numbers with respect to different features are shared, which is significantly different from previous approaches that embed features independently~\cite{gorishniy2021revisiting,somepalli2021saint,huang2020tabtransformer}.
% \end{enumerate}

\subsection{Triggers are Task Descriptions}
Incorporating task-specific information (e.g., task-prefix~\cite{zhang2022task} and textual instruction~\cite{wei2021finetuned}) can significantly improve the generalization ability of large language models, especially in few-/zero-shot scenarios. We find out that the triggers integrated into SAT-Ranker can also be viewed as task descriptions that provide meta information for search ranking. In particular, the implicit trigger can not only advise the modeling of authority and recency but also clarify the user's information need in general. For example, if a user searches for ``radioactive'', our off-the-self analysis would tell the ranking model that the major information need is ``music''. Therefore, a document entitled ``Radioactive - music by Imagine Dragon'' would be ranked higher. This opens a new gate of leveraging off-the-shelf query analysis to improve search ranking. It is very promising to employ more query analysis results as task descriptions for ranking models to boost user satisfaction modeling.
\section{Offline Experiment}
\label{sec:experiment}
\subsection{Dataset}
\label{sec:dataset}

% \begin{table}[]
% \caption{Dataset statistics.}
% \begin{tabular}{lll}\toprule
%       & {\bf \#queries} & {\bf \#documents} \\ \hline
% train & 57777     & 1731088     \\
% dev   & 14003     & 414665      \\
% test  & 6762      & 187321      \\ \bottomrule
% \end{tabular}
% \label{tab:dataset}
% \end{table}

\begin{table*}[]
\caption{Effectiveness of SAT-Ranker on overall user satisfaction and various aspects. Relative improvement or degradation with respect to the text-based ranker is reported. 
Best results are shown in {\bf bold} while second best results are underlined.
Significant difference
between text-based ranker and the corresponding method is marked with $\dagger$ ($p$ < 0.05, two-tailed paired $t$-test).}
\begin{tabular}{llcccccc} \toprule
{\bf Model}   & {\bf Feature}     & \multicolumn{2}{c}{{\bf Satisfaction}} & {\bf Quality } & {\bf Authority} & {\bf Recency } \\ 
             & & PNR           & nDCG@4           & nDCG@4  & nDCG@4   & nDCG@4       \\ \hline
SAT-Ranker$_{mlp}$ &  All  &         +0.3\% & +0.1\% & -0.3\% & -0.1\% & 0.0\% \\ \hline
\multirow{5}{*}{SAT-Ranker} & Relevance & +1.2\%*          & +0.3\%           & -1.0\%          & -0.8\%          & -0.7\%          \\
                            & Quality   & \underline{+1.6\%}*          & +0.4\%           & \underline{+3.1\%*} & +0.4\%           & +0.7\%           \\
                            & Authority & +1.0\%*          & \underline{+0.5\%}           & +1.8\%*          & \underline{+2.1\%*} & +0.9\%           \\
                            & Recency   & +1.2\%*          & +0.3\%           & -0.4\%          & -0.4\%          & \underline{+2.0\%*} 
                            \\ \cdashline{2-7}
                            
                            & \textbf{All}  & \textbf{+2.4\%*} & \textbf{+0.7\%*} & \textbf{+4.3\%*} & \textbf{+2.5\%*} & \textbf{+2.5\%*} \\ \hline
                            % \\ 
                            % \bottomrule
%SAT-Ranker    & ALL & +2.4\%$^\dagger$ & +0.7\%$^\dagger$ & +4.3\%$^\dagger$  & +2.5\%$^\dagger$  & +2.5\%$^\dagger$ \\ \bottomrule
\end{tabular}
\label{tab:overall}
\end{table*}
% We train SAT-Ranker on our manually labeled dataset.
% Queries are collected from our online platform.
% The top retrieved documents in the reranking pool of our search pipeline are labeled by expert annotators.
% We evaluate SAT-Ranker in a large-scale in-house dataset. 
We conduct offline experiments on an in-house dataset to verify the effectiveness of the proposed SAT-Ranker.
% \revision{
We collected all the training and evaluation data (i.e., query-document pairs) from the ranking stage of our search engine. The annotation is conducted on a crowdsourcing platform, where a group of annotators are trained to label each query-document pair with a satisfaction grade (i.e., from 0 to 4, indicating {bad, fair, good, excellent, perfect}). We make sure that each query-document pair is annotated by at least three annotators, and the labels are finalized by major voting when having disagreement.
% }

% \revision{
For labeling user satisfaction, we have comprehensive rules, principles and showcases made by professionals for training the annotators. Overall, the principles mainly indicate that 1) both relevance and webpage quality have major contribution to user satisfaction, such as irrelevant or low-quality results would have low grades, and 2) for authority(recency)-sensitive queries, the annotated user satisfaction would also consider the authority(recency) of the webpage. 
% }

% \revision{
For labeling each dimension of user satisfaction, such as authority, we also follow the same procedure, while the rules, principles and showcases are particularly made for the particular dimension. For example, the government websites and documents have the highest authority label and the contents published by certified professionals (e.g., doctors or lawyers) would have high authority labels if the contents are in their areas of expertise.
% }

\begin{comment}
    
To construct the evaluation dataset, we first collect a large set of queries from online search logs. For each query, we collect a set of top-ranked documents from our online search engine as the ranking candidates of the query. Each document is labeled with a 0-4 rating implying its satisfaction level.
% unsatisfying(0), partially satisfying(1), satisfying(2), highly satisfying(3).
The annotators not only consider relevance as a core judging criteria but also consider multiple aspects towards overall user satisfaction. Moreover, we further graded all the data w.r.t. quality, authority, and recency for fine-grained evaluations. During such fine-grained annotation, the annotators are required to only consider the particular dimension and ignore other dimensions of user satisfaction.
% Moreover, we further extract several subsets of data to evaluate different aspects of user satisfaction.
For example, a low-quality document could be graded as 0 w.r.t. quality, even if it is relevant to the query.
% For recency-sensitive queries, fresh documents are considered to be more satisfying than stale documents even if their relevance degree is equal, following ~\citet{yin2016ranking}.
% For authority-sensitive queries, authentic sites are considered to be more satisfying than unknown sites even if their relevance degree is equal.
%The dataset statistics is summarized in Table~\ref{tab:dataset}.

\end{comment}

\subsection{Evaluation Metrics}
\label{sec:metrics}
% Our experiment includes different metrics for offline and online evaluations.
\noindent{\bf PNR.} Positive negative ratio (PNR) measures the consistency of prediction results and ground truth. It can be formally defined as
% Formally, it's given as
\begin{equation}
    % PNR = \frac{\sum_{i,j\in [1,N]}\mathbbm{1}\{y_i > y_j\} \cdot \mathbbm{1}\{score(q, d_i) > score(q, d_j)\}}{\sum_{i,j\in [1,N]}\mathbbm{1}\{y_i > y_j\}\cdot\mathbbm{1}\{score(q, d_i) < score(q, d_j)\}}
    PNR = \frac{\sum_{i,j\in [1,n]}\mathbbm{1}\{y_i > y_j\} \cdot \mathbbm{1}\{s_i > s_j\}}{\sum_{i,j\in [1,n]}\mathbbm{1}\{y_i > y_j\}\cdot\mathbbm{1}\{s_i < s_j)\}}
\end{equation}
where $y_i$ is the ground truth label for the document $i$ and $\mathbbm{1}(\cdot)$ is an indicator function that is equal to 1 if the expression inside is true, otherwise 0.
It considers the relative ranking order of every document pair during the ranking stage.
% in the ranking list rather than focus on the top $k$ results.

\noindent{\bf nDCG@k.} Normalized Discounted Cumulative Gain (DCG) is a widely-used metric for search ranking.
It measures the ranking effectiveness discounted by document position.
Notably, it also considers document labels in the computation, which is different from PNR that only considers the partial order between labels.
% use labels only for pair generation. 
More formally, DCG is computed as
\begin{equation}
    % DCG@k = \sum_{i=1}^{k}\frac{2^{rel_{i}} - 1}{\log (1 + i)}
    DCG@k = \sum_{pos=1}^{k}\frac{2^{y_{pos}} - 1}{\log (1 + pos)}
\end{equation}
where $y_{pos}$ is the label of a document at position $pos$.
nDCG is the normalized version of DCG by dividing IDCG, which is the DCG value of an ideal ranking~\cite{DBLP:books/daglib/0021593}. We report the averaged nDCG values over all queries.
A two-tailed paired t-test is used to report the statistical significance at a 95\% confidence interval.

\begin{comment}

\noindent{\bf Med.Date@k.} For recency-sensitive queries, the median date of the top $k$ results from all the query sets is a strong indicator of how fresh these results are. We set $k=4$.

\noindent{\bf Hits@k.} For authority-sensitive queries, the hits measure whether the top results hits the target authentic documents. We set $k=1$.

\end{comment}

% To evaluate model effectiveness on quality, authority, and recency, the labels are replaced by labels on particular dimensions as $y_{pos}$, which are taken as ground truth to compute nDCG.
% Labels for document quality are divided into 3 degrees: "High quality (3)", "Fair (2)", and "Low quality (0)".  
% Labels for authority are divided into 3 degrees: "highly authentic (2)", "authentic (1)", and "non-authority-sensitive (0)".
% Labels for recency are divided into 5 degrees: "very fresh (4)", "fresh (3)", "slightly outdated (2)", "stale (1)" and "non-time-sensitive (0)".

% All the metrics are computed for each query and the averaged values over all queries are reported.

% Due to commercial issues, we only report the relative gains.

\subsection{Implementation}
\label{sec:training}
We use ERNIE-base 2.0 as the backbone~\cite{DBLP:conf/aaai/SunWLFTWW20} of SAT-Ranker.
To train the model, we adopt a multi-stage training paradigm following the previous study ~\cite{zou2021pre}.
During fine-tuning, we train SAT-Ranker on the manually labeled dataset targeting modeling user satisfaction.
In particular, we normalize and discretize all numerical features to 0-200 integers, which are subsequently tokenized in the model. Note that all the integers are included in the vocabulary of the PLM, where the pre-trained integer representations are used as the initialization for fine-tuning.
% We set the maximum sequence length to 128.
We truncate the textual input if the input length exceeds the model capacity to ensure that the numerical inputs are preserved.
We use pairwise hinge loss and a learning rate of 2e-5~\cite{zou2021pre}.
We select the best model on the dev set and report the metrics on the test set.
% Training was conducted on 8 NVIDIA A100 GPUs and took approximately 4 hours.
% After that, SAT-Ranker is adopted to rerank the top $k$ results.
% We implement all the models with PaddlePaddle \footnote{https://github.com/PaddlePaddle/Paddle}, an open-source library for developing and deploying deep learning models. 
The model is trained with 8 NVIDIA A100 GPUs, on our distributed training platform.

\subsection{Competing Methods}
\label{sec:baselines}
We compare three methods in our experiments:
\begin{itemize}[leftmargin=5mm]
\item {\bf Text-based ranker} is a vanilla PLM-based model that only uses the query text, document title, and query-aware document summary, which is originally tailored for relevance ranking.

\item {\bf SAT-Ranker$_{MLP}$} not only covers the textual inputs in the PLM backbone but also leverages a MLP network to model the numerical inputs for other aspects of user satisfaction.

\item {\bf SAT-Ranker} adopts the same inputs as SAT-Ranker$_{MLP}$ but injects the numerical inputs to the input field of the PLM.
The numerical inputs are operated in a null-prompt way~\cite{DBLP:journals/corr/abs-2203-00902}.
% All of the models are warmed up from the same checkpoint. 
\end{itemize}
In the offline experiments, we report the relative improvement of SAT-Ranker$_{MLP}$ and SAT-Ranker over text-based ranker. Note that all three models are initialized from the same checkpoint, and trained on the same training data with an identical training paradigm. The extraction of query-aware document summary is also the same, following previous study~\cite{zou2021pre}.

\subsection{Results}
\label{sec:offline_result}

% The overall evaluation results are shown in 
Table~\ref{tab:overall} shows the experimental results on all queries w.r.t. satisfaction, quality, recency, and authority. We can draw several key findings from the table as follows:
\begin{itemize}[leftmargin=5mm]
    \item SAT-Ranker$_{MLP}$ fails to achieve significant performance gain compared to vanilla text-based ranker. In particular, the improvements of SAT-Ranker$_{MLP}$ over vanilla text-based ranker are only 0.3\% and 0.1\% w.r.t. the two evaluation metrics, respectively. The results are not statistically significant. 
    The hybrid structure (i.e., SAT-Ranker$_{MLP}$) might be sub-optimal for the task of modeling user satisfaction in web search\footnote{The reason could be that the PLM module has significantly more parameters than the MLP module. When the textual and numerical features of the two modules have no in-depth interactions with each other, the PLM module would dominate the training process and the joint model (i.e., SAT-Ranker$_{MLP}$) would converge to an optimal that favors the PLM.}.
    \item SAT-Ranker significantly outperforms vanilla text-based ranker by a large margin, i.e., 2.4\% and 0.7\% on PNR and nDCG@4, respectively, in terms of overall satisfaction. This indicates that SAT-Ranker as a unified framework is preferable for modeling user satisfaction than both text-based ranker and SAT-Ranker$_{MLP}$.
    \item Despite the overall satisfaction, Table \ref{tab:overall} also demonstrates that SAT-Ranker can improve text-based ranker w.r.t. quality, authority, and recency, by a large margin. It verifies that the multi-field features are beneficial for SAT-Ranker to comprehend each dimension of user satisfaction.
    % \revision{
    % The contribution of each type of features is studied in Section~\ref{sec:ablation}.
    % }
\end{itemize}
% It can be seen that SAT-Ranker consistently outperforms ERNIE and SAT-Ranker$_{mlp}$ on all metrics tailored for different aspects.
Overall, we can conclude that both the multi-field features and the unified framework of SAT-Ranker contributes to its remarkable performance in ranking user satisfaction. Next, we analyze the contribution of each type of features in the following subsection.
% the improvement of SAT-Ranker$_{mlp}$ over text-based ranker is marginal, demonstrating that decoupling relevance and other aspects into different sub-modules is sup-optimal.
% By jointly modeling the numerical and textual inputs, 
% SAT-Ranker achieves remarkable performance, which confirms its superiority over user satisfaction. We report the relative improvement of SAT-Ranker$_{mlp}$ and SAT-Ranker over vanilla text-based ranker.
\subsection{Ablation on Separate Satisfaction Dimension}
\label{sec:ablation}
% \revision{
We conduct ablation study 
% on separate satisfaction dimension 
by adding the features on each dimension separately. 
Results are shown in Table~\ref{tab:overall}.
We draw several key findings as below:
\begin{itemize}
    \item     All the four types of features contribute significant improvements on the overall user satisfaction. Putting all together, SAT-Ranker champions on all dimensions of user satisfaction. The result confirms that the SAT-Ranker framework is a flexible and extensible framework to integrate different features that benefit the overall user satisfaction.
    \item For quality, authority and recency, we observe that adding the particular set of features is effective to improve the performance on the corresponding dimension of user satisfaction. 
    In particular, adding quality, authority, and recency features improves performance by 3.1\%, 2.1\%, and 2.0\% on their respective dimensions. Relevance features slightly improve PNR and nDCG@4, while the other dimensions of user satisfaction slightly decrease. 
    \item Quality features have the largest improvement on the PNR of user satisfaction. This is primarily because conventional text-based PLM rankers often overlook the crucial dimension of webpage quality, which significantly impacts user satisfaction.
    \item Authority features also contribute to the improvement of nDCG@4 in terms of quality. This is because authoritative webpages, such as government websites, tend to exhibit higher quality attributes, such as clear structure and presentation of content.
\end{itemize}
% }

\begin{table}[]
\caption{Effectiveness of SAT-Ranker on queries with different authority-sensitive levels. Significant difference
between text-based ranker and the corresponding method is marked with $\dagger$ ($p$ < 0.05, two-tailed paired $t$-test).}
\begin{tabular}{lcccc} \toprule
{\bf Authority Level}        & {\bf All} & {\bf Low} & {\bf Medium} & {\bf High} \\ \hline
SAT-Ranker$_{mlp}$ &-0.1\% & -0.1\% & -0.1\% & -0.2\% \\
SAT-Ranker    & +2.5\%$^\dagger$  & +2.4\%$^\dagger$  & +3.2\%$^\dagger$  & +1.8\%$^\dagger$ \\ 
\bottomrule
\end{tabular}
\label{tab:authority}
\end{table}

\begin{table}[]
\caption{Effectiveness of SAT-Ranker on queries with different recency-sensitive levels. Significant difference
between text-based ranker and the corresponding method is marked with $\dagger$ ($p$ < 0.05, two-tailed paired $t$-test).}
\begin{tabular}{lcccc} \toprule
{\bf Recency Level}        & {\bf All} & {\bf Low} & {\bf Medium} & {\bf High} \\ \hline

SAT-Ranker$_{mlp}$ & 0.0\% & 0.0\% & 0.0\% & +0.1\%  \\
SAT-Ranker    & +2.5\%$^\dagger$ & +1.8\%$^\dagger$ & +3.3\%$^\dagger$ & +4.2\%$^\dagger$  \\ \bottomrule
\end{tabular}
\label{tab:recency}
\end{table}

\subsection{User Satisfaction on Different Queries}
To further clarify the model performance on authority-/recency-sensitive queries, we divide all the queries into several groups based on the authority and recency triggers (i.e., sensitivity). In particular, all the queries are grouped into three levels, i.e., low, medium, and high.
% For authority and recency, we further report the metrics divided by query authority/recency levels in 
Tables~\ref{tab:authority} and~\ref{tab:recency} show the performances w.r.t. human-annotated user satisfaction (cf. Section~\ref{sec:dataset}) on different authority and recency levels, respectively. %where the metric is nDCG@4.
First, we can see that SAT-Ranker achieves consistent improvements for all query groups, including all levels of authority and recency. Second, the improvements become larger when the recency level is higher. This indicates that SAT-Ranker is learned to adaptively leverage the recency features to improve overall satisfaction. However, this observation is not found in the query groups divided by authority levels. In particular, the performance gain of SAT-Ranker on high authority-sensitive queries is only 1.8\%, which is lower than 3.2\% on medium authority-sensitive queries This might be because that recency is easier to be accurately modeled by the recency feature (i.e., normalized post time), while the features for modeling high authority might not be sufficient. This motivates us to incorporate more features that identify high-authority results in the future.

% From Table~\ref{tab:authority}, the Hit@1 of the highly authority demanding queries (level=2) is three times larger than the normal queries (level=0).
% By incorporating the authority signals into the model, SAT-Ranker further improves the Hit@1 accuracy of high authority demanding queries, which is not the case for normal queries.
% From Table~\ref{tab:recency}, the median date of the highly recency demanding queries (level=2) is much larger than the normal queries (level=0).
% By incorporating the recency signals into the model, SAT-Ranker consistently improves the recency of top documents across different query recency levels.

% For document quality, we use another test set that consists of 100,000 pairs of documents.
% Each document pair is semantically the same while differing in document quality including usefulness, reliability, animation, etc~\cite{moustakis2004website}.
% By incorporating document length and user dwell time, SAT-Ranker achieves 15\% improvement on the quality aspect.

\section{Online Evaluation}
\label{sec:online_result}

In this section, we deploy SAT-Ranker on the ranking system of our search engine and compare it with our online ranking system equipped with a vanilla text-based ranker. Particularly, we conduct \textbf{online interleaved comparison}~\cite{chuklin2015comparative} and \textbf{manual evaluation}.

% \subsection{System Overview}
% In general, a search system comprises retrieval and a top-$k$ reranking module.
% The top-$k$ reranking module reranks the top 40 results recalled by the retrieval module.
% SAT-Ranker is a major reranking model embedded in our top-k reranking module.
% It assigns a satisfying score to each document on receiving the features (textual and numerical), which is utilized in conjunction with other features by the top-$k$ reranking module.

\subsection{Evaluation Metrics}
We introduce three metrics for online evaluation. Interleaving is used to evaluate the online interleaved comparison, while DCG and GSB are metrics for manual evaluation. 

\noindent{\bf Interleaving.} Interleaving is an online evaluation method that comprehensively compares the strategy system with the baseline system~\cite{chuklin2015comparative}.
The results from system A and system B are interleaved and shown to the users, where the gain of the new system A over the base system
B (denoted as $\Delta_{AB}$) can be defined as
\begin{equation}
    \Delta_{AB} = 0.5\cdot\frac{wins(A) - wins(B)}{wins(A)+wins(B)+ties(A,B)}
\end{equation}
where $wins(A)$ counts how many times system A is more preferable by the users than system B, and $ties(A, B)$ counts how many times the user gives no preference over system A or B. The preference is indicated by the real online behaviors of users.

\noindent{\bf DCG.} We collect another set of queries from the online search log and manually label the ranking results of the two systems on the query set. The results are annotated with 0 to 4 grades considering user satisfaction.
% We select a pool of queries and manually label the data w.r.t. user satisfaction.
% We maintain a pool of 4000 queries from two nearest consecutive months following ~\citet{yin2016ranking}. 
% We crawl the ranked list both by the strategy and baseline systems. The unlabelled query-document pairs are collectively labeled by our expert annotators. 
We report the DCG values on top-4 ranking results.

\noindent{\bf GSB.} Good vs. Same vs. Bad (GSB), is a metric that compares two systems in a side-by-side manner.
% using user queries issued within a week.
We collect a set of queries from the online search log and ask several expert annotators to give judgments of which system should be more satisfied by the users.
% w.r.t. user satisfaction by comparing the top-ranked results from both systems. 
The front-end is similar to DiffIR~\cite{jose2021diffir}. The gain of a new system $\Delta_{GSB}$ can be formulated as:
\begin{equation}
    \Delta_{GSB} = \frac{\#Good - \#Bad}{\#Good + \#Bad + \#Same}
\end{equation}
where $\#Good$, $\#Bad$, $\#Same$ denotes the times that system A is manually judged by the annotators as better, worse, or same respectively.
We report the GSB values on top-4 ranking results.

Note that we filter out those queries with identical ranking results of the two systems during the query collection for DCG and GSB. This would prevent unnecessary annotation costs. Besides, we further extract tail queries (i.e., queries with low search frequency) from all the collected queries for performance comparison.
% at cutoff 4.

\subsection{Results}

\begin{table}[]
\caption{Relative improvements on online evaluation. }
%All the results are significant with $p<0.05$.}
\begin{tabular}{llll} \toprule
  {\bf Type} & {\bf Metric}  & {\bf All queries} & {\bf Tail queries} \\ \hline
\multirow{3}{*}{Interleaved} & $\Delta_{\mathrm{AB}}$ &  +0.77\%    &  +0.57\%    \\
&$\Delta_{\mathrm{AB-Recency}}$ &  +0.45\% &   -   \\ 
&$\Delta_{\mathrm{AB-Authority}}$ &  +0.57\% &   -   \\ \hline
\multirow{2}{*}{Manual} & $\Delta_{\mathrm{DCG}}$ &  +0.45\%    &   +0.70\%   \\
&$\Delta_{\mathrm{GSB}}$ &  +3.12\% &   +2.83\%   \\ \bottomrule
\end{tabular}
\label{tab:online_dcg_gsb}
\end{table}

% We conduct an online interleaved comparison to compare SAT-Ranker with ERNIE.
We deploy SAT-Ranker in our online ranking system to compare with our online text-based ranker.
In particular, we maintain all the other components in the online system unchanged, while only switching vanilla text-based ranker to SAT-Ranker in the experiments.
Table~\ref{tab:online_dcg_gsb} shows the online evaluation results. 
First, we find that SAT-Ranker achieves significant improvements on both all queries and tail queries, w.r.t DCG and GSB. This indicates that our proposed framework can better satisfy users.
% both on random and long-tail queries.
Second, SAT-Ranker also achieves considerable improvements in online interleaved comparison. It demonstrates that the modeling of user satisfaction can be reflected by user behaviors. To be more specific, user behaviors are more likely to occur on the top-ranked results in the new system deployed with SAT-Ranker, 
% as the users are more satisfied with the strategy system using SAT-Ranker.
Third, we observe that SAT-Ranker can also achieve online improvements for recency-sensitive and authority-sensitive queries, which is critical for satisfying diverse information needs.
% Note that we do not report recency and authority
% for recency-sensitive queries, SAT-Ranker pushes the median date of the strategy system by 5 days, which is a strong improvement in the search engine.
% For authority-sensitive queries, SAT-Ranker improves the His@1 of authentic documents by 0.57\%, which is also a strong improvement.
All the results are statistically significant (two-tailed paired $t$-test, $p$ < 0.05).

\begin{table*}[!th]
%\scalebox{0.49}{
\caption{A case on the difference between null prompt and schema prompt. Schemas are marked in \textcolor{blue}{blue}.}
\begin{tabular}{p{8cm}p{8cm}}
\toprule
{\bf Null Prompt:} & {\bf Schema Prompt:} \\ \hdashline
\begin{CJK*}{UTF8}{gbsn}
\texttt{[CLS]} 水仙花有毒可以养吗 \texttt{[SEP]}
水仙花有毒吗?可以在室内养吗? - 花百科 \texttt{[SEP]}
水仙花有毒,毒性体现在花朵、枝干以及汁液中,不会挥发到空气中,散发出来的味道是安全无毒的,因此虽然是有毒的植物,但可以养护在室内。
不过如果家里有小孩或者是宠物,养护时要注意摆放。\texttt{[SEP]}
10 5 8 10 10 \texttt{[SEP]}
0 1 2 \texttt{[SEP]}
2 2 10 5 \texttt{[SEP]}
2 3 \texttt{[SEP]}
6  2 \texttt{[SEP]}
\end{CJK*} & \begin{CJK*}{UTF8}{gbsn}
\texttt{[CLS]} 水仙花有毒可以养吗 \texttt{[SEP]}
水仙花有毒吗?可以在室内养吗? - 花百科 \texttt{[SEP]}
水仙花有毒,毒性体现在花朵、枝干以及汁液中,不会挥发到空气中,散发出来的味道是安全无毒的,因此虽然是有毒的植物,但可以养护在室内。
不过如果家里有小孩或者是宠物,养护时要注意摆放。\texttt{[SEP]}
\textcolor{blue}{词汇匹配:} 10 5 8 10 10 \texttt{[SEP]}
\textcolor{blue}{时效敏感:} 0 
\textcolor{blue}{时效:} 1 2 \texttt{[SEP]}
\textcolor{blue}{权威敏感:} 2
\textcolor{blue}{权威:} 2 10 5 \texttt{[SEP]}
%\textcolor{blue}{寻址:} 2 3 \texttt{[SEP]}
\textcolor{blue}{质量:} 6 8 \texttt{[SEP]}
\end{CJK*} 
\\
\hdashline
\texttt{[CLS]} Narcissus is poisonous, can I keep it? \texttt{[SEP]}
Is narcissus poisonous? Can I keep it indoors- Flower Encyclopedia \texttt{[SEP]}
Narcissus is poisonous. Its toxicity is reflected in flowers, branches and juice, and it will not volatilize into the air. The smell it emits is safe and non-toxic. Therefore, although it is a poisonous plant, it can be kept indoors.
However, if there are children or pets at home, be mindful of where you place things.\texttt{[SEP]}
10 5 8 10 10 \texttt{[SEP]}
0 1 2 \texttt{[SEP]}
2 2 10 5 \texttt{[SEP]}
2 3 \texttt{[SEP]}
6  2 \texttt{[SEP]}&
\texttt{[CLS]} Narcissus is poisonous, can I keep it? \texttt{[SEP]}
Is narcissus poisonous? Can I keep it indoors- Flower Encyclopedia \texttt{[SEP]}
Narcissus is poisonous. Its toxicity is reflected in flowers, branches and juice, and it will not volatilize into the air. The smell it emits is safe and non-toxic. Therefore, although it is a poisonous plant, it can be kept indoors.
However, if there are children or pets at home, be mindful of where you place things.\texttt{[SEP]}
\textcolor{blue}{lexicon matching:} 10 5 8 10 10 \texttt{[SEP]}
\textcolor{blue}{recency sensitive:} 0
\textcolor{blue}{recency:} 1 2 \texttt{[SEP]}
\textcolor{blue}{authority sensitive:} 2
\textcolor{blue}{authority:} 2 10 5 \texttt{[SEP]}
%\textcolor{blue}{navigation:} 2 3 \texttt{[SEP]}
\textcolor{blue}{quality:} 6   8 \texttt{[SEP]}
    \\ \bottomrule
\end{tabular}%
%}
\label{tab:prompt}
\end{table*}

\section{Analysis of Input Schema}
\label{sec:analysis}
Apart from the overall comparison, we also conduct investigation on what kinds of input schema lead to better performance. In particular, we mainly focus on the following research questions that we encountered in practice:
% we seek to answer the following research questions:
\begin{itemize}[leftmargin=5mm]
    \item {\bf RQ1}: Is null prompt the best for the multi-field input? Can it be further improved by schema prompt?
    \item {\bf RQ2}: What is the best way of dealing with missing values under the SAT-Ranker framework?
    \item {\bf RQ3}: How powerful is the normalized numerical expression of dates instead of using dates in a raw form?
\end{itemize}
Besides, there are many other questions that could also be investigated, as SAT-Ranker is flexible to support various forms of inputs.
% These research questions are not limited to the design choices of our SAT-Ranker framework.
% They further provide insights on the effectiveness, flexibility of the framework through the lens of prompt, language fluency, and numeracy of PLM.

\subsection{Null or Schema Prompts? (RQ1)}
\label{sec:prompt_type}

\begin{table}[]
\caption{Ranking effectiveness of SAT-Ranker using different types of prompts.}
\begin{tabular}{lll} \toprule
{\bf Prompt}          & {\bf PNR} & {\bf nDCG@4} \\ \hline
Schema        &  +2.2\%  &   +0.4\%    \\ 
Null       &  {\bf +2.4\%}   &  {\bf +0.7\% }    \\\bottomrule
\end{tabular}
\label{tab:prompt_effectiveness}
\end{table}
% \citet{DBLP:journals/corr/abs-2203-00902} empirically examined three types of prompts: template-based, schema-based, and null.
In SAT-Ranker, we apply a null prompt for the multi-field input, where no template or hint is applied to explicitly represent the meaning of each input field. Alternatively, schema prompt~\cite{DBLP:journals/corr/abs-2203-00902} can also be adopted, 
% The user satisfaction field in SAT-Ranker corresponds to the null prompt.
% Intuitively, it can also be a schema prompt so that 
where each field is explicitly expressed.
We showcase these two ways of formulating multi-field input in Table~\ref{tab:prompt}.
% The hope of the schema
The schema prompt is anticipated to provide the model with semantic meanings of the features so that those features can be better understood.
In contrast, the null prompt operates in a less explicit way, where the model must explore the hidden relationship behind these features in an end-to-end manner.
We experiment to compare the two methods.
As it can be seen in Table~\ref{tab:prompt_effectiveness}, SAT-Ranker achieves comparable ranking effectiveness in both settings while the null prompt is slightly better. More importantly, using a schema prompt is less feasible as it would significantly increase the length of the input sequence, and lead to high online latency. As such, we use a null prompt in SAT-Ranker, and the results show that 
% In summary, SAT-Ranker 
it can capture the composition between textual and numerical inputs without schema.
% The operation is computation friendly as the feature of a newly emerging aspect can be condensed into a numerical field.

\begin{comment}

\noindent{\bf Few-shot setting.} We study whether prompts contribute to different ranking effectiveness in the few-shot setting.
We report the performance using different ratios of the training data.
As can be seen from Figure x, SAT-Ranker achieves similar results when the ratio of training data is above xx.
The benefit of using schema prompts is minor.
Schema prompt yields better PNR under the few-shot setting.
Under the zero-shot setting, the schema prompt outperforms the null prompt by a larger margin, which is consistent with the recent finding that (large) language models are zero-shot reasoners~\cite{kojima2022large}.
\end{comment}

\subsection{Input Schema of Missing Values (RQ2)}

\begin{table}[]
\caption{Ranking effectiveness of SAT-Ranker using different representations for missing values.}
\begin{tabular}{lll} \toprule
{\bf Missing values}          & {\bf PNR} & {\bf nDCG@4} \\ \hline
\texttt{{[}UNK{]}} &  +2.4\%   &  {\bf  +0.7\% }     \\
%       \begin{CJK*}{UTF8}{gbsn}
% 无   \end{CJK*}
Natural language (English)      &  +2.3\%   &  +0.4\%      \\
Natural language (Chinese)
      &  {\bf +2.6\%}   &    {\bf +0.7\% }      \\\bottomrule
\end{tabular}
\label{tab:missing_value}
\end{table}

In real-world applications, missing values can occur both in offline training and online serving.
% For offline training, abandoning records with missing value comes at the cost that the losing data might be valuable.
% For online serving, the request can not be abandoned apparently.
SAT-Ranker as a practical ranking framework must learn to deal with missing values. In particular, we have two options to represent missing values in the input, i.e., 1) by natural language (i.e., "\begin{CJK*}{UTF8}{gbsn}
无  \end{CJK*}" in Chinese or ``Null'' in English), and 2) by the \texttt{{[}UNK{]}} token in the existed vocabulary.
% Under the SAT-Ranker framework, we can leverage the "unknown" knowledge of PLM.
% Such knowledge includes semantics,
% linguistic relations, and lexical features that are
% necessary to generate fluent text~\cite{nogueira2020document}.
% We explore three tokens that are semantically similar to 'missing':
% 1)Null; 
% 2)\begin{CJK*}{UTF8}{gbsn}
% 无;   \end{CJK*} 
% 3)\texttt{[UNK]}.
We conduct an experimental comparison and present the results in Table~\ref{tab:missing_value}.
As shown in the table, there is no clear winner of which way is the best, while using natural language in Chinese yields slightly better results in terms of PNR.
As our PLM backbone is trained mainly in Chinese, using an aligned token in the target language enables the model to better understand the concept of missing values. Note that we use the same natural language to represent the missing values of all features.

\subsection{Input Schema of Dates (RQ3)}
\begin{table}[]
\caption{Ranking effectiveness on the recency dimension using different forms of date.}
\begin{tabular}{lll} \toprule
{\bf Token}          & {\bf nDCG@4} \\ \hline
%w/o & 18744.3 \\ \hdashline
% YYYY-MM-DD &     18765.6        \\
% \begin{CJK*}{UTF8}{gbsn}
% YYYY年MM月DD日   \end{CJK*}
%       &       18767.2          \\ 
% Normalized  Number       &     18771.1       \\
YYYY-MM-DD &     +1.8\%        \\
% YYYY\begin{CJK*}{UTF8}{gbsn} 
% 年   \end{CJK*}
% MM\begin{CJK*}{UTF8}{gbsn} 
% 月   \end{CJK*}DD\begin{CJK*}{UTF8}{gbsn} 
% 日   \end{CJK*}
%       &       +1.9\%         \\ 
Normalized  Number       &     {\bf +2.5\%}       \\
\bottomrule
\end{tabular}
\label{tab:date_form}
\end{table}

Apart from numeracy, ~\citet{kim2022exploiting} demonstrate that PLMs capture a notion of date.
For example, the closest words to '2018' from BERT's vocabulary are "currently, October, 1990s, July,
19th"~\cite{kim2022exploiting}.
% In our implementation, 
To this end, we compare two forms of the date to inform SAT-Ranker of the recency of each document, i.e., 1) the exact date as YYYY-MM-DD, and 2) a normalized number that indicates the relative recency in the candidate documents.
% Alternatively, one can leverage the date knowledge of PLM to model recency.
% In particular, we study whether the performance in terms of recency can be improved by using two conventional forms of dates: "YYYY-MM-DD", "\begin{CJK*}{UTF8}{gbsn}
% YYYY年MM月DD日   \end{CJK*}".
As it can be seen from Table~\ref{tab:date_form}, both forms can improve the recency of the final ranking results, while
% introducing any notions of dates improves the effectiveness in recency.
% While "YYYY-MM-DD" and "\begin{CJK*}{UTF8}{gbsn}
% YYYY年MM月DD日   \end{CJK*}" encode dates in fluent natural language, they sacrifice the perception of relative recency of documents\footnote{Future works can model whole documents once so that the model perceives the dates of all documents.}.
% Remarkably, 
% our normalized date results in 
using normalized numbers has
better performance. This confirms 1) the importance of modeling recency using listwise context, and 2) the flexibility of our framework that modularizes each aspect in a numerical field.
\section{Related Work}
\label{sec:relatedwork}

% We briefly review three lines of research related to our work.

\subsection{User Satisfaction in Web Search}

%Relevance sits at the core of the system-centric evaluation paradigm~\cite{DBLP:series/synthesis/2021LinNY},
%while usefulness and satisfaction are key concepts in the user-centric evaluation of search engines~\cite{mao2016does}.
% \citet{kelly2009methods} propose a definition: {\it 'satisfaction
% can be understood as the fulfillment of a specified desire or goal'}.
Search satisfaction is defined as the fulfillment of a user’s information need~\cite{dan2016measuring}.
%By collecting data in a laboratory user study, 
~\citet{mao2016does} point out that {\it relevance} alone does not align to user satisfaction while {\it usefulness} perceived by users is a better indicator.
Similarly, ~\citet{yilmaz2014relevance} suggest that the amount of {\it effort} required to find the relevant information should be considered in conjunction with relevance.
They implied that any document that takes more time to judge than the dwell time, the time spent examining the document, is of low utility.

Several works have been proposed to improve particular dimensions of user satisfaction.
%There are, of course, a variety of aspects that contribute to different levels of user satisfaction~\cite{yin2016ranking}.
As for recency, ~\citet{dong2010towards} propose to incorporate several recency-related features including timestamp features, linktime features, webbuzz features, and page classification features.
Later, they exploited micro-blog data as a valuable source of real-time user behavior to improve recency ranking~\cite{dong2010time}.
For authority, much of the work has endeavored to link analysis that builds anchor-based web graphs~\cite{DBLP:books/daglib/0021593}, from which page authoritativeness can be obtained by graph algorithms that populate the entire graph, such as PageRank~\cite{page1999pagerank}, HITS~\cite{kleinberg1998authoritative}, and BrowseRank~\cite{liu2008browserank}.
For document quality, ~\citet{moustakis2004website} summarize nine composite criteria that correspond to quality assessment including usefulness, reliability, animation, specialization, etc.
%~\citet{hasan2011assessing} develop a quality assessment framework that covers content quality, design quality, organization quality, and user-friendly quality. 
Several works also incorporate evidence from structure information~\cite{wenniger2020structure}, gaze behaviour~\cite{mathias2018thank}, inherent visual features~\cite{shen2020multimodal,wang2020cognitive} to assess document quality.

Existing works either endeavor to the definition of user satisfaction~\cite{yilmaz2014relevance,mao2016does} or focus on the implementation of a particular dimension like authority~\cite{page1999pagerank,kleinberg1998authoritative,liu2008browserank} and freshness~\cite{dong2010towards,dong2010time}.
To the best of our knowledge, this is the first work that successfully manifests several dimensions of user satisfaction to a PLM-based ranker to improve search effectiveness.

\subsection{Pretrained Language Models for Search}

Pretrained language models, such as BERT~\cite{DBLP:journals/corr/abs-1810-04805} and T5~\cite{DBLP:journals/corr/abs-1910-10683}, have demonstrated superior performance on many natural language tasks including information retrieval~\cite{DBLP:conf/kdd/LiuLCSWCY21,zhao2022dense,ye2022fast,dong2022incorporating}.
PLMs are usually pretrained on general corpora and then finetuned on the target corpus.
For search ranking, the \texttt{[CLS]} token representation is adopted to predict whether a document is relevant to the query~\cite{DBLP:journals/corr/abs-1901-04085}.
We briefly summarize the approaches here and refer the readers to~\citet{DBLP:series/synthesis/2021LinNY} for more details.
When modeling document relevance, the evidence can be divided into sentence level~\cite{DBLP:conf/emnlp/YilmazWYZL19}, passage level~\cite{DBLP:conf/sigir/DaiC19}, and full-text level~\cite{DBLP:journals/corr/abs-2008-09093,DBLP:conf/www/WuMLZZZM20,DBLP:conf/sigir/MacAvaneyYCG19,DBLP:conf/sigir/GaoC22}.
\citet{DBLP:journals/corr/abs-2008-09093} found out that models using full-text context significantly outperform models based on sentences or passages.
Due to computing budget, modeling full-text context is intractable for web search on an industry scale.
As a trade-off, query-aware snippets or sentences are extracted as proxies to the document~\cite{zou2021pre,DBLP:journals/corr/abs-2210-08809}.
Document relevance is judged by its title and extracted sentences~\cite{zou2021pre}.
In this work, we move beyond the relevance based on texts~\cite{lin2021pretrained} and focus on the under-explored direction of modeling user satisfaction using pretrained language models.

\subsection{Numeracy in Pretrained Language Models}
Despite the fact that pretrained language models are originally designed for textual data, researches have found that pretrained token representations naturally encode numeracy and capture a prior notion of magnitude~\cite{naik2019exploring,wallace2019nlp}.
Such numeracy skill of PLM has been successfully applied on 
question answering~\cite{petrak2022improving,kim2022exploiting}, 
quantitative reasoning~\cite{liang2022mwp,lewkowycz2022solving}, 
semantic
parsing over tables~\cite{yin2020tabert,herzig2020tapas},
tabular data learning~\cite{gorishniy2021revisiting,somepalli2021saint,huang2020tabtransformer,liu2022ptab,arik2021tabnet}.
The skill can be further improved by specific digital tokenization~\cite{petrak2022improving,geva2020injecting}, task-oriented pretraining objectives~\cite{yin2020tabert,liang2022mwp,geva2020injecting}, contrastive learning~\cite{petrak2022improving}, and weakly supervised data augmentation~\cite{herzig2020tapas,geva2020injecting}.
In particular, numeracy-based PLMs~\cite{gorishniy2021revisiting,somepalli2021saint,huang2020tabtransformer,liu2022ptab,arik2021tabnet} outperform strong tree-based baselines~\cite{chen2015xgboost,ke2017lightgbm} on tabular data. However, most of the existing PLM-based methods for tabular data do not consider modeling textual fields in a unified manner. Very recently, some studies concurrent to our work propose to integrate retrieval scores as input features in language models for ranking~\cite{askari2023injecting,yu2023fusion}, while their features are simple and they do not fully investigate the effectiveness of such unified framework for modeling different types of information. In this work, the proposed SAT-Ranker is a generic method that perceives both textual and numerical inputs to comprehend different perspectives of user satisfaction, and achieves superior performance in real-world search engine.

\begin{comment}
    
question answering: 
Improving the Numerical Reasoning Skills of Pretrained Language
Exploiting Numerical-Contextual Knowledge to Improve Numerical Reasoning in Question Answering

Models
Math word problem: MWP-BERT: Numeracy-Augmented Pre-training for Math Word
quantitative reasoning: Solving Quantitative Reasoning Problems with
Language Models

most similar: context with numbers

table parsing: (row linearization) 
TABERT: Pretraining for Joint Understanding of Textual and Tabular Data
TAPAS: Weakly Supervised Table Parsing via Pre-training

multimodal: (image patch)

Revisiting Deep Learning Models for Tabular Data
SAINT: IMPROVED NEURAL NETWORKS FOR TABULAR DATA VIA ROW ATTENTION AND CONTRASTIVE
PTab: Using the Pre-trained Language Model for Modeling Tabular Data
TABPFN: A TRANSFORMER THAT SOLVES SMALL TABULAR CLASSIFICATION PROBLEMS IN A SECOND
TabTransformer: Tabular Data Modeling Using Contextual Embeddings
TabNet: Attentive Interpretable Tabular Learning

\end{comment}
\section{Conclusion}
\label{sec:conclusion}

In this paper, we focus on ranking user satisfaction in web search and formulate four critical dimensions towards user satisfaction, i.e., relevance, quality, authority, and recency. We propose a practical framework, namely SAT-Ranker, which leverages the capacities of PLMs on both textual and numerical inputs to comprehend user satisfaction. In particular, we apply a multi-field input that modularizes each dimension of user satisfaction as a set of numerical features in a unified manner. The proposed SAT-Ranker framework is extendable and flexible to be applied in real-world applications.
% Our proposed SAT-Ranker framework successfully unlocks the numeracy skill of pretrained language model, which provides a unified solution of jointly modeling all the dimensions of user satisfaction. 
We conduct rigorous offline and online experiments to demonstrate that SAT-Ranker is able to significantly improve the user satisfaction of our search engine, 
% Several design choices, such as prompt, words for dealing with missing data, and date expression are highlighted, which confirms the flexibility and extensibility of the framework.
% Overall, SAT-Ranker is effective, practical and 
and it has been fully deployed online in our ranking system.
We anticipate to motivate future research on modeling user satisfaction with pretrained language models.
% in either pretraining or fine-tuning stage.

\bibliographystyle{ACM-Reference-Format}
\bibliography{8-reference}

\end{document}